# Characteristics of Pumping Current in a YBCO Coil by a Pulse-Type Magnetic Flux Pump

Zhiming Bai, Xinhui Cui, and Chi Ma

*Abstract*—2G high temperature superconducting (HTS) wires, YBCO coated conductors, perform a better carrying current capability, which is potentially applied in the manufacture of HTS magnets. This paper presents the experimental results of the pumping current for YBCO coils using a pulse-type magnetic flux pump in the conduction-cooling system and liquid nitrogen bath (LN$_2$) cryogenic environment. Optimization of the flux pump used in the conduction-cooling system is that a constantan heater was added to keep the temperature of the pumping bridge at a certain value. Excitation effects of the YBCO coil at different temperatures were investigated in the conduction-cooling system. A fast-increasing of pumping current in the YBCO coil occurs when the temperature of the YBCO sheet (i.e., pumping bridge) is in the range of 50 K to 80 K. The relationships between saturated pumping current and input voltage, working frequency, numbers of magnetic poles were also studied. Using the seven-pole configuration, the saturated current can reach 155 A when the frequency is 20 Hz and the voltage is 6 V. The excitation characteristics of the flux pump in the LN$_2$ cooling system show the possibility of the pulse-type magnetic flux pump for the practical application of HTS magnets.

## I. Introduction

HIGH temperature superconductor (HTS) magnet has been widely applied in magnet systems such as nuclear magnetic resonance (NMR) and magnetic resonance imaging (MRI) [1][2][3]. However, HTS has an intrinsically low n-value, and the joint resistance in the HTS magnet is still relatively high. These problems will cause persistent current decay in the superconducting closed loop [4][5]. These greatly limit the applications of HTS.

Magnetic flux pump, which is subject to Faraday's law of electromagnetic induction, provides a variable magnetic flux and then injects direct current (DC) into the superconducting loop [6]. It enables superconducting coils to be excited without the direct connection of current leads, which causes great power consumption and heat leak to the cryogenic system [7]. Moreover, it can also compensate for current decay caused by HTS joints [8]. These merits make the flux pump a promising device for HTS magnet uses and triggered intensive research interest [9][10][11].

Compared with the first generation HTS wires, such as Bi-2223, the second generation HTS wires own higher operating temperature, critical current, and better on-field performance, which is more attractive for applications. YBCO coated conductor tapes with significantly improved mechanical properties show a superior behavior over bulk superconductors [12]. The development in manufacturing long length YBCO tapes brings its broad potentials in magnet working in liquid nitrogen (LN$_2$) such as MRI magnet, motor, generator, etc.

Many types of flux pumps for the YBCO coil were proposed in recent decades [13]. Coombs *et al.* presented a linear flux pump to compensate for persistent current decay in HTS magnets without electrical contact [14][15]. Recently, a persistent current switch (PCS) was proposed to charge the HTS magnet by controlling the dynamic resistance, and large dynamic resistance puts the PCS in the working state of charging mode [16]. The pulse-type magnetic flux pump has a good excitation performance in the Bi-2223 superconducting coil [17][18][19]. In this work, we use the YBCO sheet as the pumping bridge to excite the YBCO coil with the same configuration. The pumping current in YBCO coils using the flux pump in the conduction-cooling and LN$_2$ cooling system are presented. Some other characteristics of pumping current, including effects on temperature, input voltage, number of poles, and operating frequency, were also investigated and discussed in the following sections.

## II. Experimental Setup

The pulse-type magnetic flux pump is composed of four main parts: iron yoke, copper solenoids, pumping bridge, and the U-shaped frame. Eight copper solenoids wound on seven pairs of iron core to enhance the field strength that can be changed by the input voltage. The iron core is made from laminated silicon steel sheets to reduce eddy current loss. There is an air gap of 3 mm between the upper poles and the lower poles. The pumping bridge made by the YBCO coated-conductor is in the middle of the air gap. The U-shaped frame is used to support and cool the pumping bridge. More details of the pulse-type magnetic flux pump have been discussed in the former paper [17][19]. In this experiment, a constantan heater is added to the end face of the U-shaped frame. In this way, the temperature of the pumping bridge can be set at a particular value. The resistance of the constantan heater is 43.7 Ω at room temperature. Four platinum resistance thermometers are placed on the copper solenoids, the

(corresponding author: Chi Ma)
Zhiming Bai and Xinhui Cui are with the College of Science, Northeastern University, Shenyang 110819, China.

Chi Ma is with the Department of Physics, Beijing Normal University, Beijing 100875, China. (e-mail: syma_chi@163.com)

pumping bridge, the YBCO coil, and the second stage, respectively, to monitor the temperature changes in real-time.

The photograph of the experimental setup in the conduction-cooling system is shown in Fig. 1(a). The flux pump and YBCO coil are cooled by a cryocooler. The iron yoke of the flux pump is installed in the first-stage cold station, and the YBCO coil is connected to the second-stage cold station by a cooper belt. Fig. 1(b) shows the experimental setup in the $LN_2$ bath. The flux pump and coil were fixed at the bottom of the foam dewar to ensure the coil could be submerged in the liquid nitrogen bath (77 K) to keep it in the superconducting state. The superconducting current inside the YBCO coils is measured by the Hall sensor at the center, as shown in Fig. 1.

We prepared two sets of YBCO single-layer coil with the same specification, coil 1 and coil 2. They are all made of 4 mm×0.2 mm YBCO coated conductor. The critical current of the YBCO tape before winding is 220 A at 77 K (determined at 1 µV/cm, self-field). Each coil contains 5 turns. The diameter of two coils are both 110 mm and their self-inductance are 2 µH. The soldering technique used in this experiment is similar to that in Ref. [19]. The differences are that we only cleaned the oxide layer on the surface of the superconducting tape, and the solder material is replaced by Wood's metal (Bi50%-Pb26.7%-Sn13.3%-Cd10%). Wood's metal has a low melting point of 70 °C. It could avoid possible degradation in superconductivity caused by long-time high-temperature soldering. The resistance of two YBCO coils is calculated and will be discussed in the later section. Coil 1 is only used in a conduction-cooling system. However, coil 2 is used in both the conduction-cooling system and $LN_2$ system, to investigate the current pumping ability of the flux pump using different cooling methods.

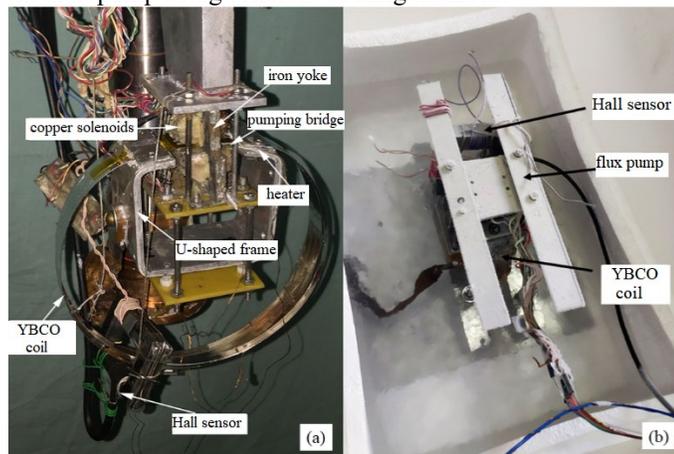

Fig. 1. Photograph of the experimental setup.
(a) conduction-cooling system (b) liquid nitrogen bath system.

When the YBCO coils had been completely cooled down, the flux pump was on and started to work. The copper solenoids were excited, and a time-varying magnetic field was generated in the gap between the upper poles and the lower poles and perpendicular to the plane of the pumping bridge. The copper solenoids were powered by a direct current supply and controlled by a control circuit [18]. Fig. 2(a) demonstrates the magnet field motion of the flux pump in this experiment. A unique winding method of the coils was applied to the flux pump in that seven magnetic poles are wrapped by eight copper solenoids, interlacing with one another. The copper solenoids are numbered by 1–8, as shown in Fig. 2(a). Eight copper solenoids were excited from left to right in a particular order. Then the superposition of the magnetic field on the magnetic pole produced a magnetic field that could move and change in amplitude. For example, when the No. 2 and the No. 3 copper solenoid are excited, the peak of the magnetic field is at the second pole, as shown by the dashed line in Fig. 2. Then the No. 3 and No. 4 copper solenoid are excited while the No. 2 copper solenoid is off. The peak of the magnetic field moves from the second pole to the third pole. The peak moves from one side of the pumping bridge to the other side in one pumping cycle. The number of the pumping cycle repeats within 1s is expressed as working frequency. The moving speed of the copper solenoids is controlled so that the working frequency of the flux pump could be adjusted.

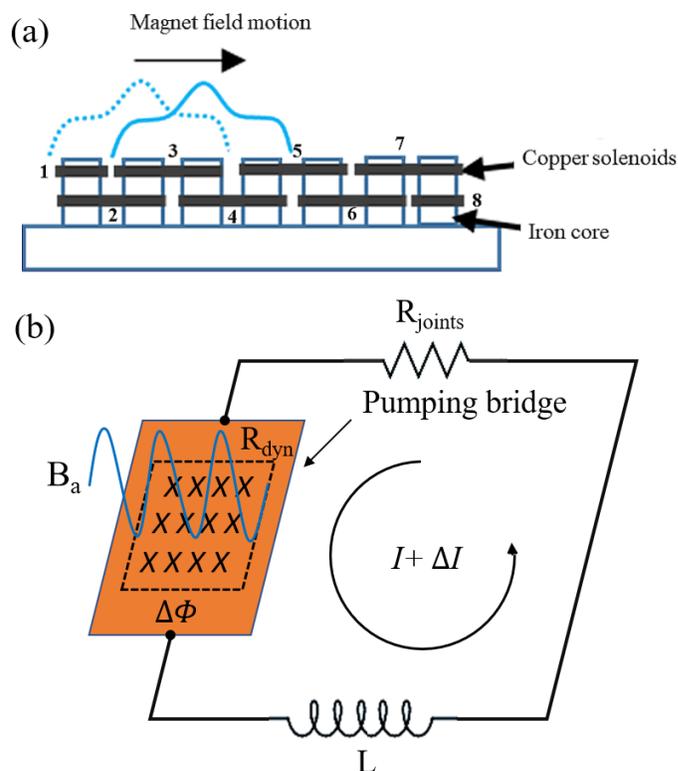

Fig. 2. The pumping mechanism of the flux pump. (a) Sketch of magnetic field motion in flux pump. (b) Circuit diagram of the superconducting closed loop.

Fig. 2(b) shows the circuit diagram of the superconducting closed loop in this experiment. $B_a$ is the time-varying magnetic field perpendicular to the pumping bridge generated by the magnetic poles in the flux pump. When the flux pump began operating and the copper solenoids are excited, $B_a$ was applied to the pumping bridge, i.e. the YBCO sheet. During each pumping cycle, the magnetic flux $\Delta\Phi$ that penetrates the YBCO sheet causes a dynamic resistance $R_{dyn}$. Meanwhile, a dynamic-resistance induced current $\Delta I$ pumping into the load coil L, i.e. YBCO coil [20]. Flux pump accumulated the induced current in every working cycle to charge the superconducting coil. When the flux pump is off, no current transports in the copper solenoids, no time-varying magnetic field will be applied to the pumping bridge, and the dynamic resistance will disappear immediately.

An expression for the dynamic resistance ($R_{dyn}$) have proposed as follow [20][21],

$$R_{dyn} = \frac{4afl}{I_{c0}}(B_a + \frac{B_a^2}{B_0})  \qquad (1)$$

Where $2a$ is the width of the superconductor sheet, $l$ is the length of the superconductor sheet, $f$ is the frequency of the applied time-varying magnetic field, $I_{c0}$ is the critical current of the tape, $B_a$ is the amplitude of the applied field, and $B_0$ is a factor describing the field-critical current relationship. The net flux injected into the superconducting loop in one working cycle is proportional to the dynamic resistance [22]. The dynamic resistance influences the increasing speed and the maximum saturated pumping current.

## III. RESULTS AND DISCUSSION

### A. Excitation effects of the temperature on the YBCO coil

First, we use the conduction-cooling system to excite the YBCO coil 1. In preceding experiments of flux pump using a Bi-2223 and $MgB_2$ sheet [18][19], the flux pump has a better working effect when the pumping bridges were operating continuously at 20 K. Like the previous process, here, the YBCO pumping bridges are also operating at 20 K, and the input voltage is 6 V, and the working frequency is 20 Hz.

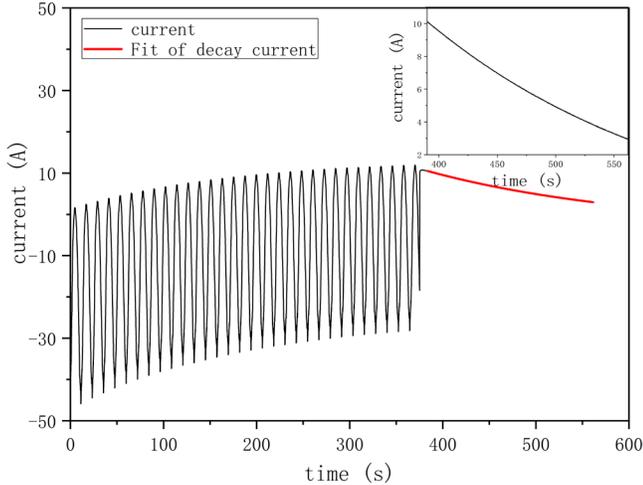

Fig. 3. The charging and decay process in the YBCO coil at 20 K. The inset shows the measured results of the decay current.

Fig. 3 demonstrates the charging and decay process of the YBCO coil at 20 K. Before the flux pump began working, the initial current in the YBCO coil is zero. A negative induced current was generated as soon as the flux pump was powered. The current jump in the YBCO coil at t=0 is caused by the initial magnetization of the iron core. Then the current in the YBCO coil increased gradually with time. When the flux pump stopped working, the current in the superconducting closed loop started to decay. This could be attributed to the resistance of two soldering joints between the pumping bridge and the YBCO coil. The decay process is indicated in red in Fig.3, and the measured results of decay current is shown in inset figure. The instantaneous value of the current can be expressed as [19]:

$$I(t) = I_0 - ke^{-\frac{t}{\tau}} \qquad (2)$$

where $I_0$ is the saturated current, $\tau$ is the decay constant and $\tau = L/R$, $L$ is the inductance of the YBCO coil, $R$ is the joint resistance. We use equation (2) to fit the decay current to evaluate the joint resistance. The fitting results indicate that the decay constant $\tau$ is about 224 s, and the inductance is about 2 µH, thus the joint resistance of coil 1 is calculated as 0.009 µΩ. Using the same method, the resistance of coil 2 is calculated as 0.01 µΩ.

As shown in Fig. 3, the current in the YBCO coil fluctuated wildly. It can be concluded that only a small net flux could be trigged into the superconducting closed loop when the flux pump was working in one cycle at 20 K. The maximum current reached 12 A after the flux pump continuously working 370 s. The average current increasing rate at 20 K is only 0.03 A/s. To speed up the current increase, it is necessary to get greater dynamic resistance. From equation (1), we know that the dynamic resistance depends on the frequency, the critical current of the tape, and the amplitude of the applied field. In the following experiments, we will regulate these three parameters to improve the performance of the flux pump. Compared with Bi-2223 and $MgB_2$ sheets used in the former experiment, this YBCO coated conductor has a higher critical current at the same temperature and magnetic field strength. The magnetic flux accumulated in the YBCO superconducting loop each cycle is relatively small, and the current increase in the coil is small. The critical current of the YBCO tape will reduce when its temperature increase, and then the dynamic resistance increases. So, we change the temperature to study the excitation effects.

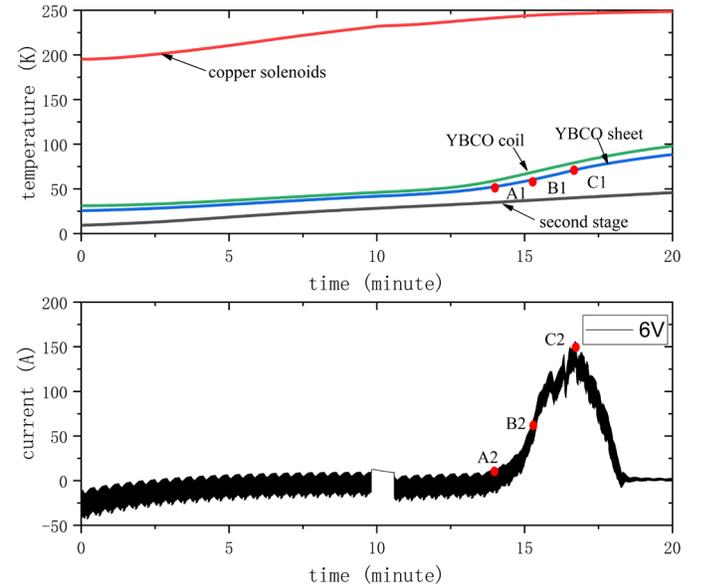

Fig. 4. Measured results of pumping current and its corresponding temperature. (a) The changing curve of temperature with time. The red, green, blue, and black lines represent the temperature of copper solenoids, YBCO coil, YBCO sheet, and the second-stage cold station, respectively. (b) Current measured in YBCO coil as a function of time for various temperatures.

Four platinum resistance thermometers are placed on the copper solenoids, the YBCO sheet, the YBCO coil, and the second-stage cold station separately. When the temperature of the second-stage cold station arrives at about 20 K, we switch off the cryocooler and switch on the flux pump with the input voltage of 6 V, and measure the temperature and the pumping current. The temperature in the cryocooler system rose because copper solenoids were excited and generated Ohmic heat. Fig.4 shows the measured results of current and its corresponding temperature, in which the measured temperature profiles are shown in Fig. 4(a), and Fig. 4(b) shows the curve of pumping current over time.

As shown in Fig. 4(a), the temperature of the second-stage cold station, the flux pump, and the YBCO loop increased gradually. The entire cryogenic system started to heat up. The temperature of the YBCO sheet at 14 mins is 53 K and marked as point A1. The pumping current at 14 mins is 10 A and marked as point A2. The average current increasing rate before 14mins is small. When the temperature of the YBCO sheet is higher than 53 K, the increasing speed of the pumping current has a significant rise. To the point B2 marked in Fig. 4(b), the increasing speed of pumping current is the largest, and the corresponding temperature marked as B1 in Fig. 4(a) is 56 K. The fast-increasing trend of pumping current indicates the dynamic resistance increases when the temperature rises and the net magnetic flux pumped into the superconducting loop in a single cycle increases. This trend stopped and then the pumping current reached the maximum value at point C2. The corresponding temperature in Fig. 4(a) is 71 K and marked as point C1. After then, as the temperature increased, although the flux pump continued working, the current in the coil decreased. This current decay could be mainly attributed to the relatively high and unstable temperature environment, and the magnetic field imposed on the YBCO sheet makes quench in the superconducting loop.

TABLE I
MEASURED RESULTS OF PUMPING CURRENT IN DIFFERENT INPUT VOLTAGE WITH VARYING TEMPERATURE

| Input voltage (V) | Maximum current (A) | The temperature of the YBCO sheet(K) |
| --- | --- | --- |
| 3 | 49 | 76 |
| 5 | 144 | 66 |
| 6 | 150 | 71 |
| 7 | 156 | 75 |

Experiments are also performed when the input voltage is 3 V, 5 V, and 7 V. The results of the maximum pumping current and the corresponding temperature of the YBCO sheet are shown in Table 1. When the input voltage was set as 3 V, 5 V, 6 V, and 7 V, maximum currents are 49 A, 144 A, 150 A, and 156 A, respectively. The temperatures of the YBCO sheet with four different input voltages are in close range, 76 K, 66 K, 71 K, and 75 K.

The fast-increasing part appeared when the temperature of the pumping bridge is in the range of 50 K to 80 K. This is close to the temperature range of liquid nitrogen. When considering practical applications, another set of flux pump working in the $LN_2$ bath was prepared to observe the effect of current pumping and discussed in later sections.

*B. Excitation effects of the input voltage on the YBCO coil*

The magnetic field strength affects the magnetic flux pumped in each pumping cycle, and it depends on the input voltage of the flux pump. In this part of work, we have designed two sets of cooling systems: a conduction-cooling system and a $LN_2$ bath system. Coil 2 was used to do experiments in those two cases to explore the excitation effects of the input voltage. In the conduction-cooling system, we use the constantan heater to keep the temperature of the YBCO coil near 75 K so that it could be taken as a comparison test with the $LN_2$ system. The working frequency is set as 20 Hz. Fig. 5 shows the results of the pumping current with the different input voltage in the conduction-cooling system.

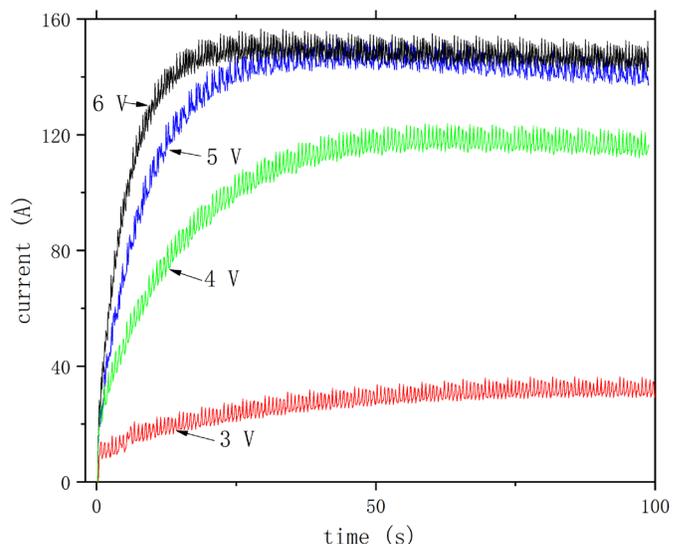

Fig. 5. The pumping current with different input voltages in the conduction-cooling system.

Current curves with different input voltage have different slopes and saturated values. When the input voltage was 6 V, the current in the YBCO coil reached 155 A after the flux pump working for 24 s. The average current increasing speed in this situation is 6.5 A/s. However, when the input voltage was 3 V, the current only increased to 24 A after the flux pump working for the same 24 s. Increasing the input voltage could improve the current pumping ability of flux pump. Compared with the experiment results of 20 K in Fig. 3, the temperature rise makes the average current increasing speed increase by 217 times. It is obvious that the increasing speed of pumping current in the YBCO coil is significantly improved with the increase of the operating temperature.

Fig. 6 shows the results of the saturated pumping current with the different input voltage in the conduction-cooling and $LN_2$ system. It can be seen from Fig. 6 that two curves follow the same trend. The saturated pumping current is very small when the input voltage is 2 V. The magnetic field imposed on the pumping bridge is not strong enough, thus resulting in only a little magnetic flux injecting into the YBCO sheet. The saturated current grows significantly when the input voltage increases from 2 V to 6 V. The magnetic field strength increases obviously with the input voltage, and more magnetic flux begins to pump into the pumping bridge. The pumping current can reach 155 A in the conduction-cooling system and 153 A in the LN$_2$ system when the input voltage is 6 V. However, it nearly has no apparent increase from 6 V to 10 V. From the relationship between the saturated pumping current and input voltage, it can be concluded that the high-performance input voltage of the flux pump is between 4 V and 6 V. And the best working voltage for the flux pump in this experiment is 6 V.

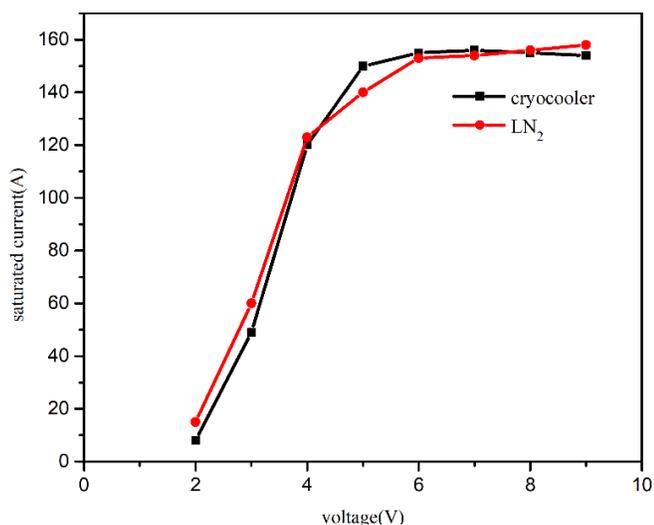

Fig. 6. Measured results of saturated current with different input voltages in conduction-cooling and LN$_2$ system.

*C. Excitation effects of the frequency on the YBCO coil*

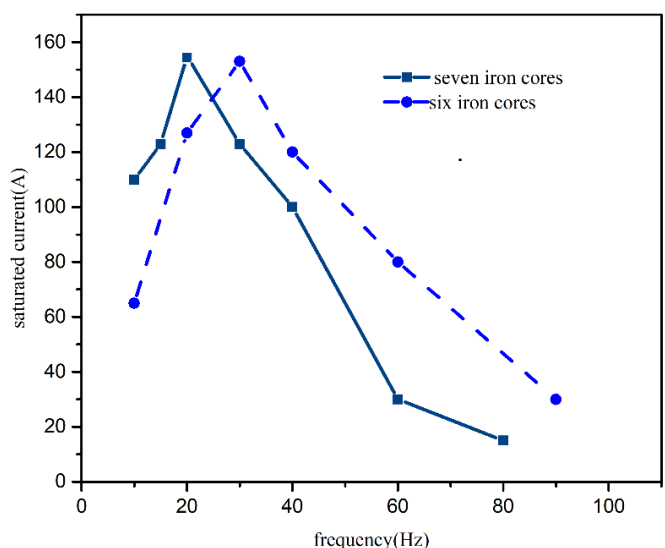

Fig. 7. Measured results of the saturated current with different working frequencies and numbers of magnetic poles in the LN$_2$ system.

The moving speed of the magnetic field generated by the magnetic pole is also a factor affecting the pumping current. The working voltage of the flux pump was set as 6 V, and the saturated pumping current at different working frequencies was studied in the LN$_2$ system. In this part, only six poles were excited to investigate the effects of the number of poles on maximum saturated pumping current. The experimental results of the six-pole configuration were compared with the original seven-pole one. Fig. 7 shows the relationship between the saturated current and the working frequency in the six- and seven-pole configurations. The maximum saturated current began to increase with the increase of working frequency. According to equation (1), the dynamic resistance is proportional to the working frequency. A large dynamic resistance will induce a large current in the coil during each working cycle of the flux pump. The seven-pole and six-pole configurations reach the maximum saturated current with a frequency of 20 Hz and 30 Hz, and the maximum saturated current is 153 A and 152 A, respectively. However, as the working frequency continues to increase, the maximum saturated current decreases instead.

The magnetic poles are made of copper solenoids wound on laminated silicon steel sheets. In the process of generating a moving magnetic field by the flux pump, the increase of working frequency may cause more loss in iron cores. The core loss ($\Delta P$) includes two parts: hysteresis loss ($\Delta P_h$) and eddy current loss ($\Delta P_e$). It can be expressed as:

$$\Delta P = \Delta P_h + \Delta P_e = Af + Bf^2 \qquad (3)$$

where $f$ is the frequency, A and B are constants determined by the material factors of the core. $\Delta P_h$ and $\Delta P_e$ are both related to frequency. The relationship between the core loss and frequency shows a parabolic growth. We have taken some measures to reduce the core loss, such as using the soft magnetic material silicon steel with a narrow hysteresis loop as the iron core to reduce the hysteresis loss and stacking the silicon steel sheet to reduce the eddy current loss. However, when the frequency is high, the core loss cannot be ignored. The core loss generates heat, which causes a temperature rise, leading to local demagnetization in magnetic poles. The effective magnetic flux pumped in each pumping cycle decreases and, consequently, the maximum saturated pumping current in the YBCO coil decrease. Other possible reasons will be verified and analyzed in future experiments.

## IV. CONCLUSION

The YBCO coils are prepared and excited by the pulse-type magnetic flux pump in the conduction-cooling and LN$_2$ system. A constantan heater is added to the U-shaped frame of the flux pump to change the temperature of the pumping bridge. The excitation effects of the flux pump at different temperatures have been investigated in the conduction-cooling system. This HTS flux pump device applies the dynamic resistance mechanism, which uses a time-varying magnetic to trigger flux flow into the superconducting closed loop. When the temperature of the YBCO sheet is in the range of 50 K to 80 K, the pumping current in the YBCO coil shows a fast-increasing trend. The characteristic of pumping current at different input voltage and frequency has been investigated. The YBCO coil could obtain a saturated current of 155 A when the input voltage of the flux pump is 6 V and the pumping frequency is 20 Hz with seven magnetic poles. For the six-pole configuration, the pumping current has the 152 A maximum value at 30 Hz. The experiment results in the LN$_2$ system show the pulse-type magnetic flux pump is workable for the practical applications of HTS magnets.